\documentclass[12pt]{iopart}

\usepackage{iopams}
\usepackage{graphicx}% Include figure files

\begin{document}

\title{Quantum dots and spin qubits in graphene}
\author{Patrik Recher and Bj{\"o}rn Trauzettel}

\address{Institut f{\"u}r Theoretische Physik und Astrophysik, University of W\"urzburg, 97074 W\"urzburg, Germany}
\ead{precher@physik.uni-wuerzburg.de, trauzettel@physik.uni-wuerzburg.de}
\begin{abstract}
This is a review on graphene quantum dots and their use as a host for spin qubits. We discuss the 
advantages but also the challenges to use graphene quantum dots for spin qubits as compared to the more standard
materials like GaAs. We start with an overview of this young and fascinating field and will then discuss gate-tunable quantum dots in detail. We calculate the bound states for three different quantum dot architectures where a bulk gap allows for confinement via electrostatic fields: (i) graphene nanoribbons with armchair boundary, (ii) a disc in single-layer graphene, and (iii) a disc in bilayer graphene. In order for graphene quantum dots to be useful in the context of spin qubits,  one needs to find reliable ways to break the valley-degeneracy. This is achieved here, either by a specific termination of graphene in (i) or in (ii) and (iii) by a magnetic field, without the need of a specific boundary. 
%These gate-tunable  quantum dots in the long-run might provide the host for the ideal spin qubit in graphene quantum dots. 
We further discuss how to manipulate spin in these quantum dots and explain the mechanism of spin decoherence and relaxation caused by spin-orbit interaction in combination with electron-phonon coupling, and by hyperfine interaction with the nuclear spin system.
\end{abstract}

\maketitle
%%%%%%%%%%%%%%%%%%%%%%%%%%%%%%
%%%%%%%%%%%%%%%%%%%%%%%%%%%%%%
\section{Introduction}
%%%%%%%%%%%%%%%%%%%%%%%%%%%%%%
\label{intro}
%%%%%%%%%%%%%%%%%%%%%%%%%%%%%%
Quantum dots are tiny islands of electrons where charging effects as well as quantum confinement play a crucial role \cite{Kouwe1997}. Besides fundamental insights, these artificial atoms could also work as building blocks for the control of electronics at the single electron level. So called single-electron transistors are able to switch on and off electron transport through a dot by means of electrical gates using the effect of Coulomb blockade. The electron not only has a charge but also spin---its intrinsic angular momentum $s=1/2$. The ability to use the spin rather than the charge of the electron to control electrical conduction is used in spintronics \cite{Wolf2001}, a quite recent and very active field in mesoscopic physics. Exploiting the spin degree of freedom, a quantum dot can act as a spin filter \cite{Reche2000,Folk2003,Hanso2004} or as a spin-blockade device \cite{Ono2002}. Spin being a quantum mechanical variable, can also be controlled in a coherent manner. Indeed, quantum dots have been proposed as a host for a quantum bit (qubit) where the spin of an electron residing in the quantum dot forms the fundamental two-level system of a qubit \cite{Loss1998}. Arrays of such quantum dots with tunable tunnel-couplings between them would work as a universal quantum computer \cite{Cerle2005}. Recent progress in this field using GaAs-based two-dimensional electron gases is impressive \cite{Hanso2007} but some rather limiting sources of spin decoherence have been identified and attributed to the rather strong spin-orbit interaction and to the hyperfine interaction with the nuclear-spin system of the host material.

In this review, we would like to give an overview on the field of quantum dots in graphene---a single sheet of graphite \cite {Geim2007,Castr2009}. The first part of the review focuses on our recent theoretical work on quantum dots in single-layer and bilayer graphene that can be controlled and formed by electrical gates. These proposals solve two main problems of graphene quantum dots as building blocks for electron spin-based quantum computing: The challenge of achieving a controllable confinement and the reliable breaking of the valley degeneracy. The second part of the review is devoted to spin manipulation and decoherence in graphene quantum dots. Here, we put an emphasis on the two most prominent spin decoherence channels: spin-orbit interaction in combination with electron-phonon coupling to lattice vibrations and hyperfine interaction with the surrounding nuclei.

The review is organized as follows: We start with a general overview of the field of graphene quantum dots in Section \ref{overview}. In Section \ref{sec_dots}, we will discuss the bound states as a function of confinement and/or as a function of magnetic field in three different geometries: (i) graphene nanoribbons, (ii) discs in single-layer graphene, and (iii) discs in bilayer graphene. The three proposals (i)-(iii) have in common that electrical gates can define the spatial extend of the quantum dot, however the way the valley degeneracy of graphene is broken---a prerequisite for spin qubits in graphene---is different for the three proposals. In Section \ref{spinqubits}, we present and discuss the physics of spin qubits in graphene quantum dots including an extended discussion of spin decoherence properties. Finally, in Section 5, we conclude and give a brief outlook.
%%%%%%%%%%%%%%%%%%%%%%%%%%%%%%%%%%
%%%%%%%%%%%%%%%%%%%%%%%%%%%%%%%%%%
\section{General overview of graphene quantum dots}
\label{overview}
%%%%%%%%%%%%%%%%%%%%%%%%%%%%%%%%%%
%%%%%%%%%%%%%%%%%%%%%%%%%%%%%%%%%%
Graphene is the first truly two-dimensional condensed matter system discovered experimentally in the group of A. Geim at the University of Manchester in 2004 \cite{Novos2004}. The low-energy dynamics of electrons is governed by states near two inequivalent $K$-points in the Brillouin zone ---so called valleys---and resembles the one of massless Dirac fermions \cite{Semen1984,Divin1984,Fradk1986,Halda1988} where a new pseudospin degree of freedom appears due to the two sublattices defining the honeycomb lattice of graphene.  This low-energy Dirac Hamiltonian indeed describes the physics of electrons and holes in the vicinity of the Dirac points as experimentally confirmed by the existence of the half-integer quantum Hall effect in graphene \cite{Novos2005,Zhang2005}.

Quantum dots in graphene have been extensively discussed in the recent literature -- both from the theory side as well as from the experimental side.
Theoretically, many approaches to graphene quantum dots exist: Graphene islands can be created by cutting flakes of graphene into the desired shape. In this case, boundary conditions and chaotic behavior become relevant \cite{Raedt2009,Wurm2009}. On the other hand, alternative approaches exist for electrostatically defined quantum dots, including quasi bound states in single-layer \cite{Silve2007,Chen2007,Matul2008,Barda2009} and bilayer graphene \cite{Matul2008} as well as true bound states in single-layer \cite{Trauz2007,Reche2009,Schne2008} and bilayer graphene \cite{Perei2007,Reche2009,Perei2009}. An electrostatic potential alone cannot produce true bound states unless some very specific symmetries are met \cite{Barda2009} because of the phenomenon of Klein tunneling \cite{Katsn2006,Cheia2006}. To overcome this problem, one needs to find a way to generate a gap in the bulk of graphene. Another possibility to confine Dirac fermions is by using spatially inhomogeneous magnetic fields \cite{Marti2007}. Furthermore, quantum antidots, that can be created in graphene by cutting out holes, have been proposed as potential host systems for spin qubits \cite{Peder2008}.

Experimentally, graphene quantum dots are created by etching or scratching graphene islands attached to leads via constrictions. Single electron transistor behavior \cite{Stamp2008,Ponom2008} as well as quantum confinement \cite{Ponom2008,Schne2009} have been observed, more recently also in double quantum dots \cite{Molit2009,Molit2010}. Energy levels have also been studied in magnetic fields where indication of the formation of Landau levels was observed \cite{Schne2009,Guetti2009}. The zeroth Landau level then can be used to locate the electron-hole crossover in graphene quantum dots \cite{Guetti2009}. Observation of quantum confinement as well as Coulomb blockade effects in electron transport requires that an electron is trapped in the dot long enough such that size quantization and charging energy effects become relevant. The constrictions separating the dot from the open contacts act as tunnel barriers. It is believed that the constrictions are tiny nanoribbons where a gap is formed in graphene due to confinement. Graphene nanoribbons in combination with a back gate and top gates themselves can be used to form single quantum dots \cite{Liu2009} as well as double quantum dots \cite{Liu2009b}.
To understand the behavior of the level structure in edged graphene quantum dots and graphene nanoribbons, edge disorder as well as Coulomb blockade effects play a crucial role \cite{Sols2008,Wunsc2008,Haeusl2009}. Due to the large disorder induced by the edges it is rather difficult to control the level-structure of the quantum dots. Especially in graphene nanoribbons, disorder plays a dominant role \cite{Stamp2009,Oosti2010}.
Nevertheless, new experiments report the observation of excited states \cite{Schne2009} as well as spin states \cite{Guetti2010}, pointing towards some improvements in control and stability of graphene quantum dots.

We mention in passing that it has been theoretically shown that edge states, prominent in graphene nanoribbons with zigzag edges, are present in any quantum dot shape with sharp boundaries \cite{Akhme2008,Wimme2010} which seemingly predicts a rather universal feature of graphene nanoflakes.

%%%%%%%%%%%%%%%%%%%%%%%%%%%%%%%%%%%
%%%%%%%%%%%%%%%%%%%%%%%%%%%%%%%%%%%
\section{Electrostatically defined quantum dots in graphene}
\label{sec_dots}
%%%%%%%%%%%%%%%%%%%%%%%%%%%%%%%%%%%
%%%%%%%%%%%%%%%%%%%%%%%%%%%%%%%%%%%
In this section,  we discuss our approach to use electric fields to confine electrons in
graphene and bilayer graphene. A straightforward approach as used in GaAs quantum dots \cite{Hanso2007}
is not possible due to Klein tunneling. Several ways are possible to induce a gap in bulk graphene. In general, quantum confinement can lead to the opening of a gap in ribbons \cite{Nakad1996,Brey2006,Son2006}. Within the tight-binding approximation of graphene, armchair boundary conditions can lead to an insulator and gate-tunable quantum dots in ribbons can be created. Another promising direction is to start with bulk graphene and induce a gap via the interaction with a substrate \cite{Giova2007,Zhou2007,Ender2010}.
In the following, we discuss three quantum dot architectures that allow for bound states in graphene tunable by electrostatic fields: (i) graphene nanoribbons with armchair-terminated boundaries, (ii) discs in single-layer graphene, and (iii) discs in bilayer graphene. Special emphasis is given on the ability to controllably break the valley degeneracy, a prerequisite for two-qubit gates \cite{Trauz2007,Reche2009} for spin-based quantum computing in graphene.
%%%%%%%%%%%%%%%%%%%%%%%%%%%%%%%%%%%%%%%%%%%%%%%%%%%%%%%%%%%%%%
%%%%%%%%%%%     Graphene Nanoribbons %%%%%%%%%%%%%%%%%%%%%%%%%%%%%%%%%
%%%%%%%%%%%%%%%%%%%%%%%%%%%%%%%%%%%%%%%%%%%%%%%%%%%
\subsection{Quantum dots in graphene nanoribbons}
\begin{figure}[t]
\begin{center}
\includegraphics[width=6cm]{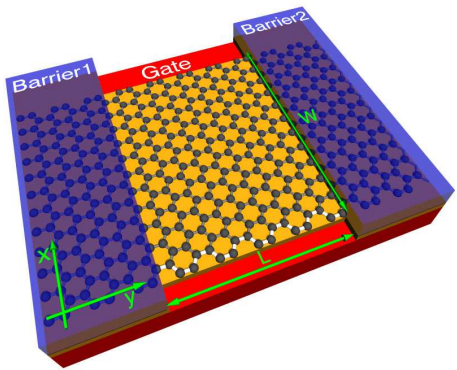}
\end{center}
\caption{{\bf Quantum dot in graphene nanoribbon.}  A ribbon of graphene with semi-conducting armchair
boundaries is schematically shown. Two barrier gates (blue) define
the rectangular size of the quantum dot (with width $W$ and length
$L$). A back gate (red) allows one to shift the energy levels in
the dot. Two or more quantum dots of this type can be easily put in series in a single nanoribbon as will be discussed in more detail below (after Ref.~\cite{Trauz2008}).} \label{setup}
\end{figure}

We first concentrate on a single quantum dot which is assumed to
be rectangular with width $W$ and length $L$, see
Fig.~\ref{setup}.
%%%%%%%%%%%%%%%%%%%%%%%%%%%%%%%%%%%%%%%
%%%%%%%%%%%%%%%%%%%%%%%%%%%%%%%%%%%%%%%
%%%%%      FIGURE    1          %%%%%%
%%%%%%%%%%%%%%%%%%%%%%%%%%%%%%%%%%%%%%%
%%%%%%%%%%%%%%%%%%%%%%%%%%%%%%%%%%%%%%%
The basic idea of forming the dot is to take a ribbon of graphene
with semiconducting armchair boundary conditions in $x$-direction
and to electrically confine particles in $y$-direction.

The low energy properties of electrons (with energy $\varepsilon$
with respect to the Dirac point) in such a setup are described by
the $4\times4$ Dirac equation
\begin{equation} \label{Dirac}
\frac{\hbar v}{i} \left(\begin{array}{cc}
\sigma_{x}\partial_{x}+\sigma_{y}\partial_{y}&0\\
0&-\sigma_{x}\partial_{x}+\sigma_{y}\partial_{y}
\end{array}\right)\Psi+\mu(y)\Psi =\varepsilon\Psi,
\end{equation}
where the electric gate potential is assumed to vary stepwise,
$\mu(y) = \mu_{\rm gate}$ in the dot region (where $0\le y\le L$),
and $\mu(y) = \mu_{\rm barrier}$ in the barrier region (where
$y<0$ or $y>L$). In Eq.~(\ref{Dirac}), $\sigma_x$ and $\sigma_y$
are Pauli matrices (denoting the sublattices in graphene).
The four component spinor envelope wave
function $\Psi =
(\Psi^{(K)}_A,\Psi^{(K)}_B,-\Psi^{(K')}_A,-\Psi^{(K')}_B)$ varies
on scales large compared to the lattice spacing. Here, $A$ and $B$
refer to the two sublattices in the two-dimensional honeycomb
lattice of carbon atoms, whereas $K$ and $K'$ refer to the vectors
in reciprocal space corresponding to the two valleys in the
bandstructure of graphene.
The appropriate semiconducting armchair boundary conditions for
such a wave function can be written as ($\alpha = A,B$)
\cite{Brey2006}
\begin{eqnarray} \label{bc}
\Psi_\alpha^{(K)}|_{x=0} &=& \Psi_\alpha^{(K')}|_{x=0} , \nonumber
\\
\Psi_\alpha^{(K)}|_{x=W} &=& e^{\pm 2\pi/3}
\Psi_\alpha^{(K')}|_{x=W} .
\end{eqnarray}
These boundary conditions couple the two valleys and are, thus,
the reason why the valley degeneracy is lifted \cite{foot1}. It is
well known that the boundary condition (\ref{bc}) yields the
following quantization conditions for the wave vector $k_x \equiv
q_n$ in $x$-direction \cite{Brey2006,Tworz2006}
\begin{equation} \label{qn1}
q_n = (n \pm 1/3)\pi/W , \;\; n \in \mathbb{Z} .
\end{equation}
The level spacing of the modes (\ref{qn1}) can be estimated as
$\Delta \varepsilon \approx \hbar v \pi/3W$, which gives $\Delta
\varepsilon \sim  30 \,{\rm meV}$, where we used that $v
\sim 10^6 \, {\rm
 m/s}$ and assumed a quantum dot width of about $W \sim 30 \, {\rm nm}$.
Note that Eq.~(\ref{qn1}) also determines the energy gap for
excitations as $E_{\rm gap} = 2 \hbar v q_0$. Therefore, this gap
is of the order of 60 meV, which is unusually small for
semiconductors. This is a specific feature of graphene that will
allow for long-distance coupling of spin qubits as will be
discussed in Sec.~\ref{spinqubits} below.

\begin{figure}[t]
\vspace*{0.5cm}
\begin{center}
\includegraphics[width=6.5cm]{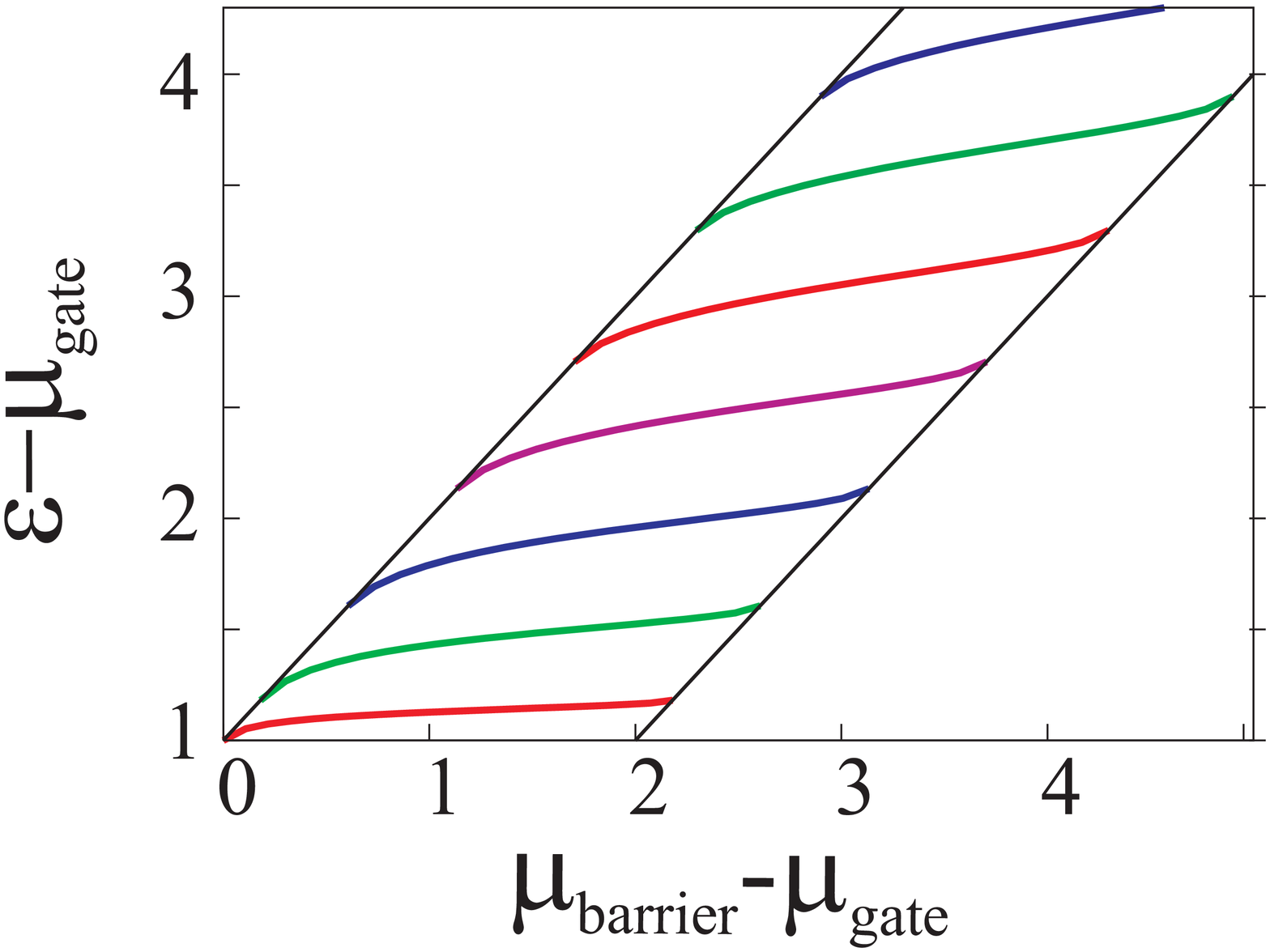}
\caption{\label{fig2} {\bf Quantum dot levels.} Bound-state solutions of a dot with aspect
ratio $q_0 L=\pi L/3W=5$. The diagonal lines indicate the region
in which bound-state solutions do exist given by
Eq.~(\ref{window1}). All energies are taken in units of $\hbar v
q_0$ (after Ref.~\cite{Trauz2008}).}
\end{center}
\end{figure}

We now present in more detail the ground-state solutions, i.e.
$n=0$ in Eq.~(\ref{qn1}). The corresponding ground-state energy
$\varepsilon$ can be expressed relative to the potential barrier
$\mu=\mu_{\rm barrier}$ in the regions $y<0$ and $y>L$ as
$\varepsilon=\mu_{\rm barrier} \pm \hbar v
(q_{0}^{2}+k^{2})^{1/2}$.
Here, the $\pm$ sign refers to a conduction band ($+$) and a
valence band ($-$) solution to Eq.~(\ref{Dirac}).
For bound states to exist and to decay at $y \rightarrow \pm
\infty$, we require that $\hbar v q_0 > |\varepsilon -\mu_{\rm
barrier}|$, which implies that the wave vector $k_y \equiv k$ in
$y$-direction, given by
\begin{equation} \label{kbarrier}
k = i \sqrt{q_0^2 - ((\varepsilon -\mu_{\rm barrier})/\hbar v)^2}
,
\end{equation}
is purely imaginary.
In the dot region ($0 \leq y \leq L$), the wave vector $k$ in
$y$-direction is replaced by $\tilde{k}$, satisfying
$\varepsilon=\mu_{\mathrm{gate}}\pm \hbar v
(q_0^{2}+\tilde{k}^{2})^{1/2}.$ Again the $\pm$ sign refers to
conduction and valence band solutions. (In the following, we focus
on conduction band solutions to the problem.)
In the energy window
\begin{equation} \label{window1}
|\varepsilon-\mu_{\rm gate}| \ge \hbar v q_0 >
|\varepsilon-\mu_{\rm barrier}| ,
\end{equation}
the bound state energies are given by the solutions of the
transcendental equation
\begin{equation} \label{trans1}
\tan(\tilde{k} L) = \frac{\hbar v \tilde{k} \sqrt{(\hbar v q_0)^2
- (\varepsilon -\mu_{\rm barrier})^2}}{(\varepsilon - \mu_{\rm
 barrier})(\varepsilon - \mu_{\rm gate}) - (\hbar v q_0)^2} .
\end{equation}
We show a set of solutions to Eq.~(\ref{trans1}) for a dot with
aspect ratio $q_0 L=\pi L/3W=5$ in Fig.~\ref{fig2}.

%%%%%%%%%%%%%%%%%%%%%%%%%%%%%%%%%%%%%%%
%%%%%%%%%%%%%%%%%%%%%%%%%%%%%%%%%%%%%%%
%%%%%      FIGURE    2          %%%%%%
%%%%%%%%%%%%%%%%%%%%%%%%%%%%%%%%%%%%%%%
%%%%%%%%%%%%%%%%%%%%%%%%%%%%%%%%%%%%%%%
%%%%%%%%%%%%%%%%%%%%%%%%%%%%%%%%%%%%%%%%%%%%%%%%
%%%%%%%%%%%%%%   Disc single layer     %%%%%%%%%%%%%%%%%%%%%%%
%%%%%%%%%%%%%%%%%%%%%%%%%%%%%%%%%%%%%%%%%%%%%%%%
\subsection{Graphene disc in single-layer graphene}
\label{singledisc}
%%%%%%%%%%%%%%%%%%%%%%%%%%%%%%%%%%%%%%%%%%%%%%%%
In this subsection, we discuss a quantum dot in {\it gapped} graphene defined via electrostatic gates without
requiring a certain boundary condition as the quantum dot will be purely defined electrostatically by a potential $U(x,y)$.
The interaction with a substrate can induce different potentials for the two sublattices thereby breaking inversion symmetry  \cite{Giova2007,Zhou2007,Ender2010}. This interaction gives rise to a
mass term $M\sigma_z$ in the Dirac equation. Then the Hamiltonian in the valley-isotropic form \cite{Beena2008,remark1} and with the inclusion of a perpendicular homogeneous magnetic field is \cite{Reche2009}
%%%%%%%%%%%%%%%%%%%%%%%%%%%%%%%%%%%%%%%%%%%%%%%%%%%%%%%%%%%%
\begin{equation} \label{htau}
H_\tau = H_0 + \tau \Delta \sigma_z + U(x,y),
\end{equation}
%%%%%%%%%%%%%%%%%%%%%%%%%%%%%%%%%%%%%%%%%%%%%%%%%%%%%%%%%%%%
where $H_0 = v({\bf p} + e{\bf A}) \cdot {\bf \sigma}$, ${\bf B} =
\nabla \times { \bf A} = (0,0,B)$, $v=10^6$ m/s is the Fermi velocity and $\tau = \pm$ differentiates the
two valleys $K$ and $K'$. We choose the symmetric gauge ${\bf A} =\frac{B}{2} (-y,x,0)$ and assume a circular symmetry in the
confinement potential $U(x,y)=U(r)$ with $r=\sqrt{x^2 + y^2}$. In the following we set $\hbar=1$.
%
%%%%%%%%%%%%%%%%%%%%%%%%%%%%%%%%%%%%%%%
%%%%%%%%%%%%%%%%%%%%%%%%%%%%%%%%%%%%%%%
%%%%%      FIGURE    Setup and dot levels       %%%%%%
%%%%%%%%%%%%%%%%%%%%%%%%%%%%%%%%%%%%%%%
%%%%%%%%%%%%%%%%%%%%%%%%%%%%%%%%%%%%%%%
\begin{figure}
\vspace{0.5cm}
\begin{center}
\hbox{\qquad\qquad\resizebox{5.8cm}{!}{\includegraphics{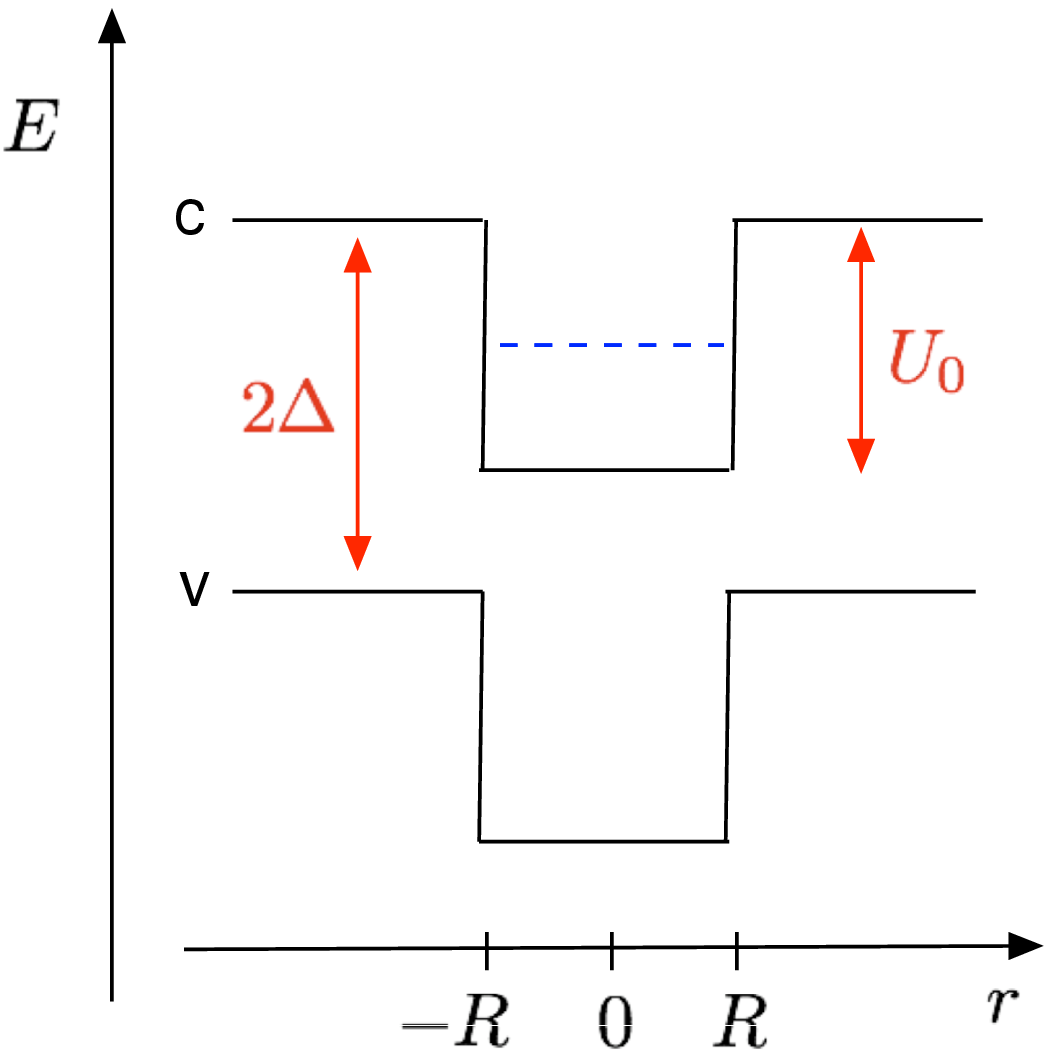}}\qquad\resizebox{6.0cm}{!}{\includegraphics{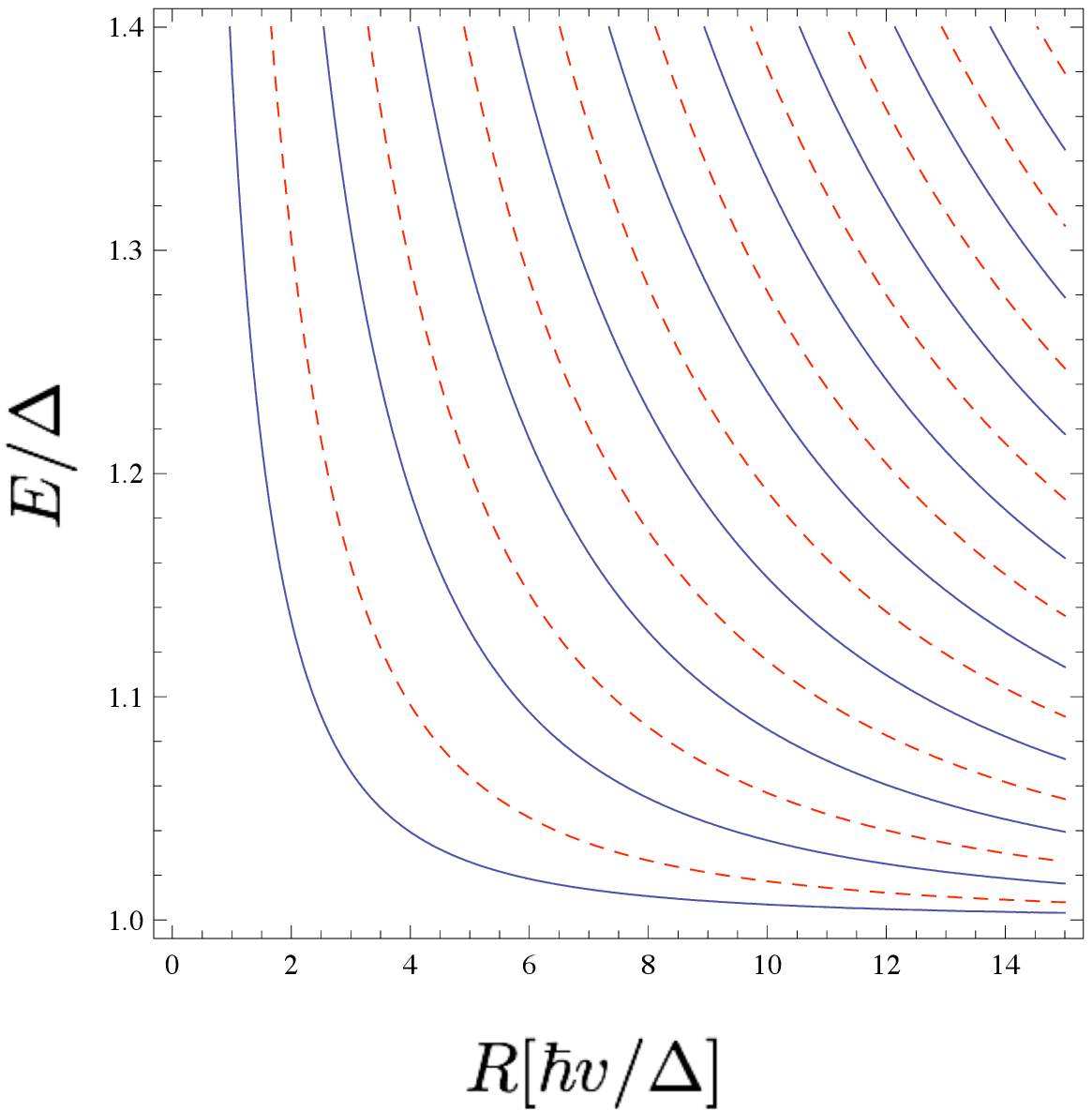}}}
\end{center}
\caption{\label{Setupdisc}
{\bf Left: Energy diagram for quantum dot in single-layer graphene.} A constant mass term $\Delta$ and an electrostatic potential with height $U_{0}$ give rise to bound states (dashed line) in the conduction band (c) defining a quantum dot of radius $R$. Note that the confining potential $U(r)$ is repulsive for holes in the valence band (v). \\
{\bf Right: Bound state levels as function of dot radius.} Energy levels for $U_{0}=\Delta$ and for $j=1/2$ at zero magnetic field. Full lines correspond to $\tau=+1$, dashed lines to $\tau=-1$ (after Ref.~\cite{Reche2009}).}
\end{figure}
%%%%%%%%%%%%%%%%%%%%%%%%%%%%%%%%%%%%%%%%

Since $H_\tau$ commutes with the total angular momentum operator $J_{z}=-i\partial_{\varphi}+\sigma_{z}/2$, the energy eigenspinors can be chosen to be eigenstates of $J_{z}$
\begin{equation}
\Psi^\tau(r,\varphi) = e^{i(j-1/2) \varphi} \left( \begin{array}{c}
\chi_A^\tau(r) \\ \chi_B^\tau (r) e^{i \varphi} \end{array}
\right) ,
\end{equation}
with $j$ the eigenvalue of $J_{z}$ which has to be an half-odd integer.
To find the bound states of Eq.~(\ref{htau}) we solve the radial Dirac equation
$
\tilde{H}_\tau(r) \chi^\tau(r) = E \chi^\tau(r),
$
with $\chi^\tau(r)=(\chi_A^\tau(r),\chi_B^\tau(r))^{T}$ and
\begin{equation}
\tilde{H}_\tau (r) = -iv \sigma_x \partial_r + \tau \Delta
\sigma_z + U(r) + \\v \sigma_y  \left(
\begin{array}{cc} \frac{j-1/2}{r} + br & 0 \\ 0 &
\frac{j+1/2}{r} + br \end{array} \right) ,
\end{equation}
where $b\equiv eB/2$.
The radial-part of the spinor solutions for bound states inside ($U(r)=0$) and outside ($U(r)=U_{0}$) the quantum dot are expressed in terms of hypergeometric functions $M(a,b,z)$ $(r<R)$ and $U(a,b,z)$ $(r>R)$ \cite{Reche2009}, where $R$ is the dot radius, see Fig.~\ref{Setupdisc}.

Matching the solutions at $r=R$, we find the characteristic equation for the energy eigenvalues $E$ of quantum dot bound states.
For $j>0$, we obtain
\begin{eqnarray}
\label{eq:E1}
\xi_{>}^{+}\,M(q_<,j+1/2,x)\,U(q_{>},j+3/2,x)&&\nonumber\\\qquad\qquad-\xi_{<}^{+}\,M(q_{<},j+3/2,x)\,U(q_{>},j+1/2,x)=0,&&
\end{eqnarray}
and for $j<0$ we obtain
\begin{eqnarray}
\label{eq:E2}
\xi_{>}^{-}\,
M(q_<,-j+3/2,x)\,U(q_{>}-1,-j+1/2,x)&&\nonumber\\\qquad\qquad-\xi_{<}^{-}\,M(q_{<}-1,-j+1/2,x)\,U(q_{>},-j+3/2,x)=0,&&
\end{eqnarray}
where $x\equiv bR^2$=$(1/2)(R/l_{B})^2$ with $l_{B}=\sqrt{\hbar/eB}$ the magnetic length. Without loss of generality, we choose here $B$ positive. The bound state levels for negative $B$ are obtained from the symmetry ${\tilde H}_{\tau}(j,B)={\tilde H}_{-\tau}(-j,-B)$. We have further introduced the parameters $q_{<,>}=(j-1/2)\,\theta(j)+1-(\epsilon_{<,>}^2-\Delta^2)/4bv^2$, $\xi_{<}^{+}=(\epsilon_{<}-\tau\Delta)/4(j+1/2)$, $\xi_{>}^{+}=bv^2/(\epsilon_{>}+\tau\Delta)$, $\xi^{-}_{<}=(j-1/2)/(\epsilon_{<}+\tau\Delta)$ and $\xi_{>}^{-}=1/(\epsilon_{>}+\tau\Delta)$ with $\theta(x)$ the Heaviside step function, and $\epsilon_{<}\equiv E$, $\epsilon_{>}\equiv E-U_{0}$.
%%%%%%%%%%%%%%%%%%%%%%%%%%%%%%%%%%%%%%%%%%%%%%%%%%%%%%%%%%%
%%%%%%%%%%%%%%%%%%%%%%%%%%%%%%%%%%%%%%%%%%%%%%%%%%%%%%%%%%%
%%%%%%%%%  Figure finite B-filed single-layer dot %%%%%%%%%%%%%%%%%%%%%%%%%%%
%%%%%%%%%%%%%%%%%%%%%%%%%%%%%%%%%%%%%%%%%%%%%%%%%%
%%%%%%%%%%%%%%%%%%%%%%%%%%%%%%%%%%%%%%%%%%%%%%%%%%%%%%%%
\begin{figure}[h]
\label{singlelayerfig1}
\vspace{0.4cm}
\begin{center}
\includegraphics[width=0.5\columnwidth]{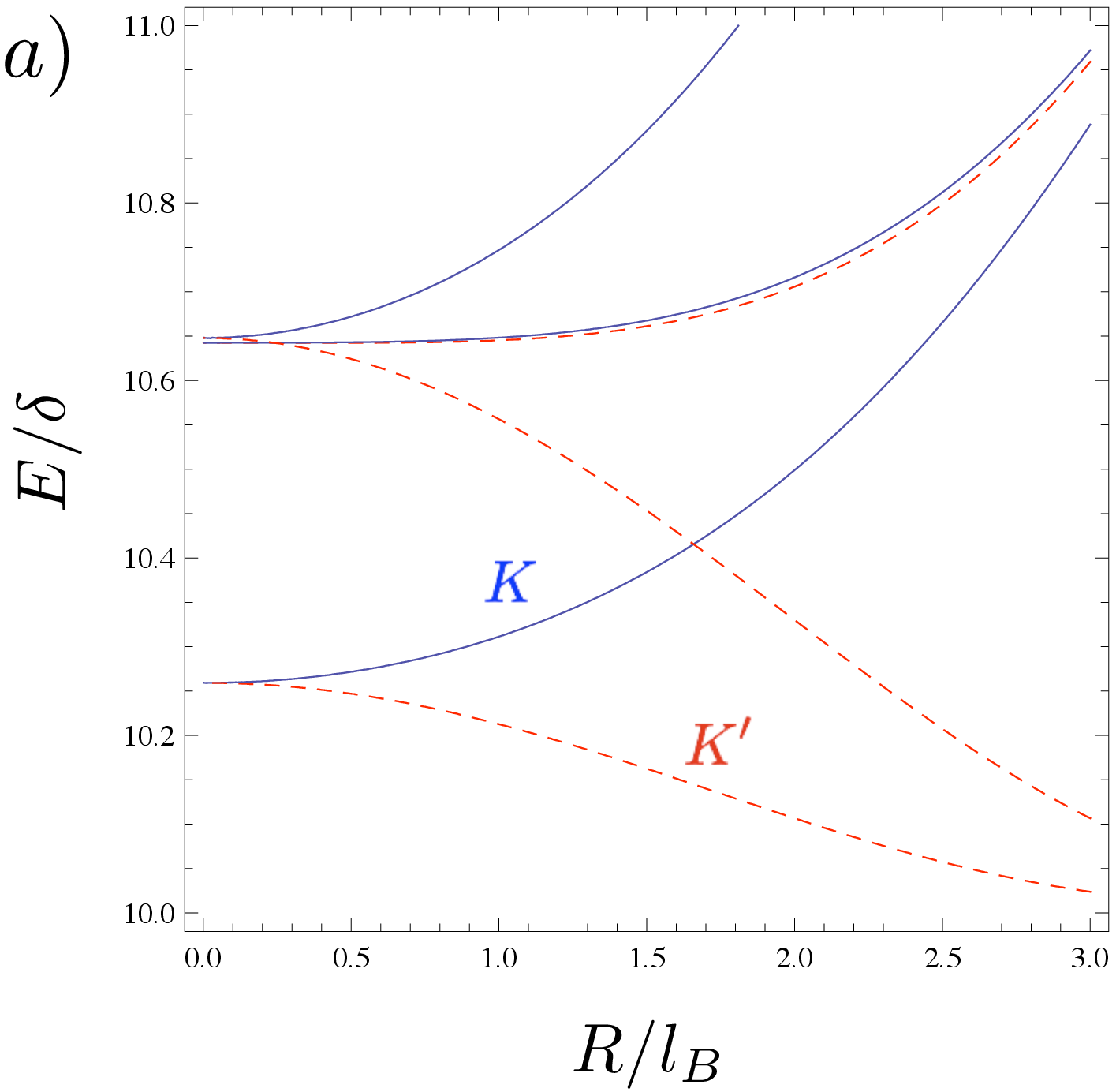}
\end{center}
\begin{center}
\includegraphics[width=0.5\columnwidth]{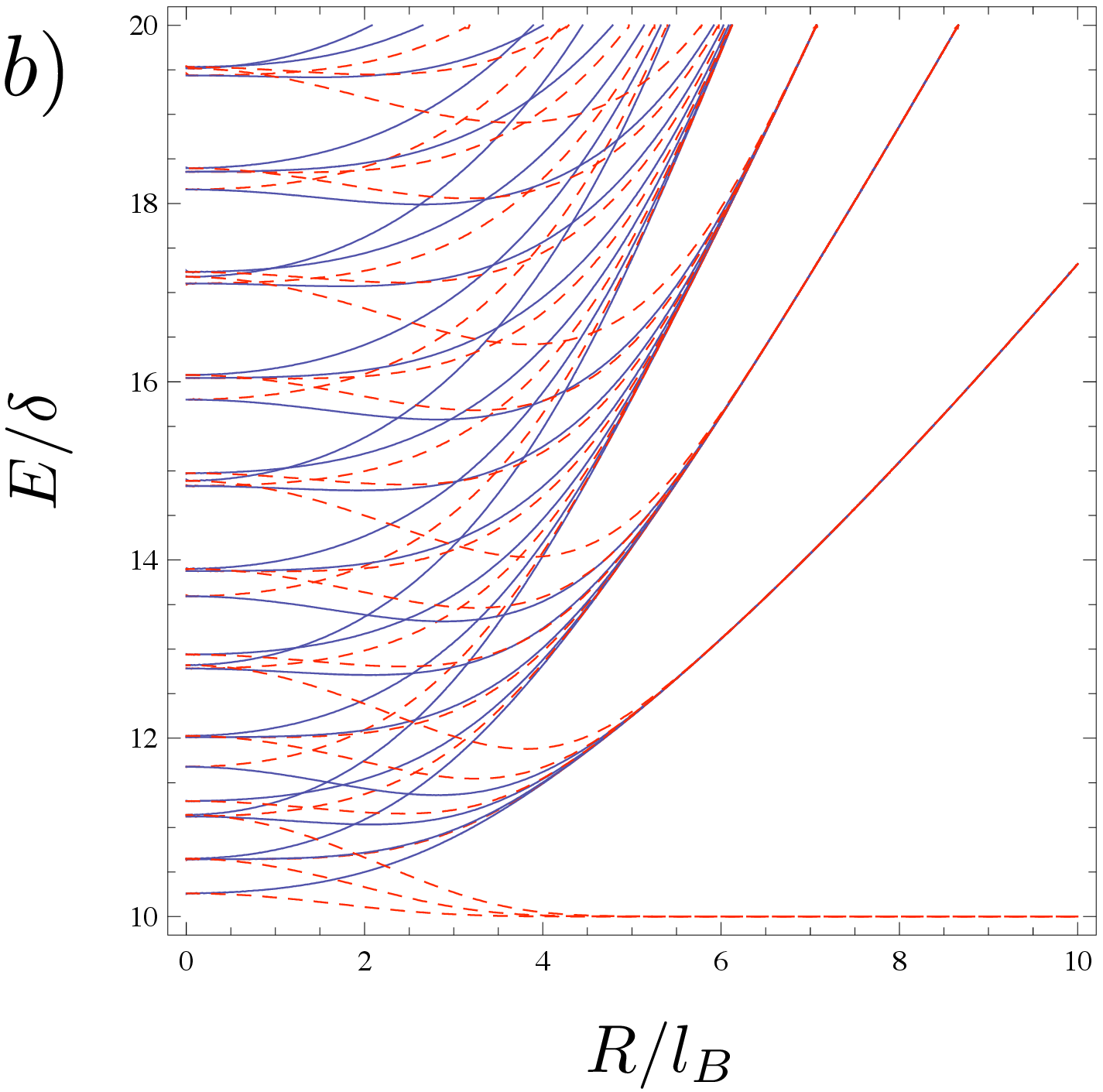}
\end{center}
\vspace{-0.4cm}
\caption{{\label{Fig22}}{\bf  Dot levels at finite magnetic fields.} a) Numerical evaluation of characteristic equations (\ref{eq:E1}) and (\ref{eq:E2}) as a function of $R/l_{B}$ with $l_{B}=(\hbar/eB)^{1/2}$ the magnetic length and $R$ the dot radius. We use $\Delta=10\,\delta$ with $\delta = \hbar v/R$ and $U_{0}=\Delta$. a) At  small $B$-fields we observe a breaking of the level degeneracy, shown for $(j,\tau)=(\frac{1}{2},\pm1),(-\frac{1}{2},\pm 1)$, and $(\pm\frac{3}{2},\pm 1)$. The full lines are for $\tau=1$ and
dashed lines are for $\tau=-1$ corresponding to the two valleys of graphene. b) Same parameters as in a),  but for larger magnetic fields. The energy levels converge to the bulk Landau levels with increasing $R/l_{B}$. Included are levels for $j=\pm\frac{1}{2},\pm\frac{3}{2}$, and $\pm\frac{5}{2}$ (after Ref.~\cite{Reche2009}). }
\label{singlelayerfig}
\end{figure}

In Fig.~\ref{Setupdisc} (right panel), we show the evolution of quantum dot levels as a function of dot size for angular-momentum quantum number $j=1/2$. Full lines and dashed lines
correspond to the two valleys. Due to the symmetry $E(j,\tau)=
E(-j,-\tau)$, the two set of curves display also the cases $j=1/2$ and
$j=-1/2$ in the same valley. The different solutions for the dashed
and full lines are therefore a direct consequence of effective time-reversal symmetry (eTRS) breaking in a
single valley at zero magnetic field by a finite mass term \cite{Berry1987}, see Sec. \ref{symmetries}.

In Fig.~\ref{singlelayerfig}, we show the bound states of the quantum dot as a function of magnetic field evaluating the characteristic equations Eqs.~(\ref{eq:E1}) and (\ref{eq:E2}) numerically.
In Fig.~\ref{singlelayerfig}(a), we show the low-lying bound states in the conduction band. Note that the valley-degeneracy (or orbital degeneracy) is broken at finite magnetic field. The largest level spacing between the (non-degenerate) groundstate and first excited state we estimate from Fig.~\ref{singlelayerfig}(a) to be at $R/l_{B}\sim 1.8$ and is about 165 meV/$R$[nm] for the parameters used in Fig.~\ref{singlelayerfig}. At $R/l_{B}\sim 1.8$, we obtain for the Zeeman splitting $\Delta_z=g\mu_{B}B\sim 200$ meV/$R^2$[nm] using $g=2$ which shows that the level spacing is always larger than the Zeeman energy for reasonable dot sizes. Considering a QD with $R=25$ nm, we obtain a valley splitting $\Delta_{KK'}$ at $R/l_{B}\sim 1.8$ of about 6.6 meV corresponding to 77 K, being much larger than $4$ K, the temperature achieved by cooling with liquid helium. The necessary magnetic field corresponding to $R/l_{B}= 1.8$ is $B$=3.41 T (and $B=0.85$ T for $R=50$ nm with $\Delta_{KK'}\sim 3.3$ meV)  which is also easily achievable in the laboratory. A gap of size 0.23 eV has been concluded from ARPES data in graphene on top of a SiC substrate \cite{Zhou2007}. Therefore, the gap $\Delta$ and also the confining potential step height $U_{0}$ could easily be larger than the QD level spacing $\delta = \hbar v/R$ which is about 26 meV. These results suggest that such QDs confined in graphene could be an ideal host for spin qubits where the orbital degeneracy is controllable by a magnetic field. Note that these quantum dots will not suffer from edge roughness as much as quantum dots defined in graphene nanoribbons.

In Fig.~\ref{singlelayerfig}(b), we show the merging of the quantum dot states with the bulk Landau levels (LLs)
\begin{equation}
E_{n}=\pm\delta\sqrt{(\Delta/\delta)^2+2n(R/l_{B})^2};\,\,n=1,2,3,...
\end{equation}
with increasing magnetic field. Note in particular, that there is a zero mode LL at $E=-\tau\Delta$ which lies entirely in one valley \cite{Halda1988}. In contrast, in Ref.~\cite{Schne2008}, the zero mode LL is at $E=0$ as a result of the assumption of a finite mass outside the dot, but zero mass inside the dot.

%%%%%%%%%%%%%%%%%%%%%%%%%%%%%%%%%%%%%%%%%%%%%%%%%%%%%%%%%%%
%%%%%%%%%%%%%%%%%%%%%%%%%%%%%%%%%%%%%%%%%%%%%%%%%%%%%%%%%%%
\subsection{Graphene disc in bilayer graphene}
\label{bilayerdisc}
\begin{figure}[h]
\vspace{0.4cm}
\begin{center}
\includegraphics[width=0.5\columnwidth]{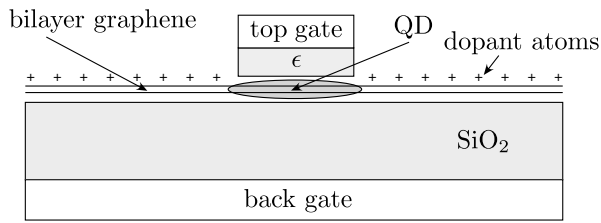}
\end{center}
\vspace{-0.4cm}
\caption{{\bf Quantum dot in bilayer graphene.} A back gate and dopants on top of the bilayer control the voltage $V$ between the layers---leading to a controllable gap opening---as well as the Fermi energy (band filling). An additional top gate allows to induce a spatially inhomogeneous electrostatic potential $U(r)$ analogous to the single-layer model which leads to bound states in the conduction band (or valence band) of the bilayer. Another possibility is to use a split top gate (instead of a combination of top gate and dopants) to achieve a similar confinement (after Ref.~\cite{Reche2009}).}
\label{layout}
\end{figure}

%%%%%%%%%%%%%%%%%%%%%%%%%%%%%%%%%%%%%%%%%%%%%%%%%%%%%%%%%%%
%%%%%%%%%%%%%%%%%%%%%%%%%%%%%%%%%%%%%%%%%%%%%%%%%%%%%%%%%%%
Bilayer graphene is the two-layer analog of the single layer, which in addition is coupled by a tunneling matrix element
$t_\perp$ coupling two sublattices of different layers (the so-called Bernal stacking). A voltage $V$ between the two layers breaks inversion symmetry (like the mass term $\Delta$ in the single layer) and opens a gap proportional to the voltage \cite{McCan2006,Min2007}. In addition, the combination of a top gate and a back gate allows to tune the gap and the average potential $U(r)$ independently, see Fig.~\ref{layout}.
%%%%%%%%%%%%%%%%%%%%%%%%%%%%
%%%%%%%%%%%%%%%%%%%%%%%%%%%%
%%%Setup Fig bilayer%%%%%%%%%%%%%%%%
%%%%%%%%%%%%%%%%%%%%%%%%%%%

To investigate gate-tunable quantum dots in bilayer graphene, we separate the Hamiltonian in the bilayer into two parts: ${\cal H} = {\cal H}_0 + {\cal H}_1^{\tau}$. ${\cal H}_0$ encodes the motion of the electrons within the planes and is given by two copies of the Dirac equation. In the valley-isotropic representation it takes on the form ($\hbar=v=1$)
\begin{equation}
\label{eq:H0b_1}
{\cal H}_0 =\left(
\begin{array}{cccc}
0 & p_x + i p_y & 0 & 0 \\
p_x - i p_y & 0 & 0 & 0 \\
0 & 0 &  0 & p_x - i p_y  \\
0 & 0 & p_x + i p_y & 0
\end{array}\right),
\end{equation}
in both valleys. Like in the case of the single layer we add a magnetic field by the minimal coupling prescription ${\bf p}\rightarrow( {\bf p}+e{\bf A})$ with ${\bf A}=(B/2)(-y,x,0)$.
The other part of the Hamiltonian (i.e. ${\cal H}_1^{\tau}$) encodes the biasing field and
the hopping $t_{\perp}$ between the two planes. The interplane hopping matrix element $t_\perp$ has recently been measured to be $t_\perp = 0.40$ eV \cite{Zhang2008,Li2009}. In the simplest approximation,
we may take
\begin{equation}
\label{eq:H1b_1}
{\cal H}_1^{\tau} =\left(
\begin{array}{cccc}
\frac{\tau V}{2} & 0 & t_\perp & 0 \\
0 & \frac{\tau V }{2} & 0 & 0 \\
t_\perp & 0 &  -\frac{ \tau V}{2} & 0  \\
0 & 0 & 0 & -\frac{\tau V}{2}
\end{array}\right) + U(r) {\bf 1},
\end{equation}
with $U(r)$ the applied electrostatic potential profile as defined in the previous section. The index $\tau = \pm 1$ again distinguishes the two valleys (note that in the valley-isotropic representation the basis is chosen such that the two planes in the bilayer are exchanged in the spinors that describe different valleys). The same Hamiltonian ${\cal H}$ has been analyzed in Ref.~\cite{Perei2007} at zero magnetic field and in Ref.~\cite{Perei2009} at finite magnetic field. However, the confinement described in Eq.~(\ref{eq:H1b_1}) by $U(r)$ has been achieved in Refs.~\cite{Perei2007, Perei2009} by a position dependent "mass term" $V(r)$, instead, leading to qualitative and quantitative differences as the gap is absent in the dot (i.e. $V=0$ in the dot center), and electrons as well as holes can be confined in the dot, simultaneously

To diagonalize ${\cal H}$ (i.e. to find the eigenspinors $\Psi$ that fulfill ${\cal H }\Psi = E \Psi$) we go to cylindrical coordinates in which the states are easily classified according to their conserved value of total angular momentum $m$ ($m$ being an integer). More explicitly, we factor out the angular dependence of the states according to
\begin{equation}
\label{eq:transform1}
\Psi =
\frac{e^{ i m \varphi}}{\sqrt{r}}
\left(\begin{array}{cccc}
1 & 0 & 0 & 0 \\
0& e^{- i \varphi} & 0 & 0 \\
0& 0 & 1 & 0 \\
0 & 0 & 0 & e^{ i \varphi}
\end{array}\right)
\Psi_1.
\end{equation}
Note that the angular momentum in the bilayer case is an integer $m$, in contrast to the half-odd integer $j$ in the single layer case, which reflects the different pseudospins in the bilayer (pseudospin 1) and single-layer (pseudospin 1/2).
With the definitions $j = m+1/2$ and $s={\rm sgn}(B)$, the Hamiltonian ${\cal H}_0$, which now acts on  $\Psi_1$, can be written as
\begin{eqnarray}
\label{eq:H0b_trans1}
{\cal H}_0 =
\frac{1}{i \sqrt{2} l_B }&&\nonumber\\
\times\left(\begin{array}{cccc}
0 & \partial_{\xi} - \frac{j-1}{ \xi} -s \xi & 0 & 0 \\
\partial_{\xi} +\frac{j-1}{ \xi} +s \xi & 0 & 0 & 0 \\
0 & 0 &  0 & \partial_{\xi} +\frac{ j}{ \xi} +s \xi  \\
0 & 0 & \partial_{\xi} - \frac{j}{ \xi} -s \xi & 0
\end{array}\right).&&
\end{eqnarray}
In the latter equation, we have introduced the dimensionless coordinate $\xi = r /(\sqrt{2} l_B)$, where $l_B = \sqrt{\hbar / (e |B|)}$ is the magnetic length.
The eigenvalue problem can now be solved by using the general
solutions of the ordinary differential equation imposed by ${\cal H}_0$.
The details of finding the eigenvectors and eigenenergies are somewhat more involved as in the single layer case, and we refer the interested reader to Ref.~\cite{Reche2009}. The eigenspinors are again expressed in terms of hypergeometric functions $M(a,b,z)$ (inside the dot) and $U(a,b,z)$ (outside the dot). For a fixed energy $E$, two solutions for constant $U(r)$ exist (corresponding to the two bands of the bilayer at positive $E$), so that the bound-state solutions are found by matching linear combinations of pairs of solutions inside and outside the dot at $r=R$ which requires to find the determinant of a 4$\times$4 matrix.

\begin{figure}[h]
\vspace{0.4cm}
\begin{center}
\includegraphics[width=0.9\columnwidth]{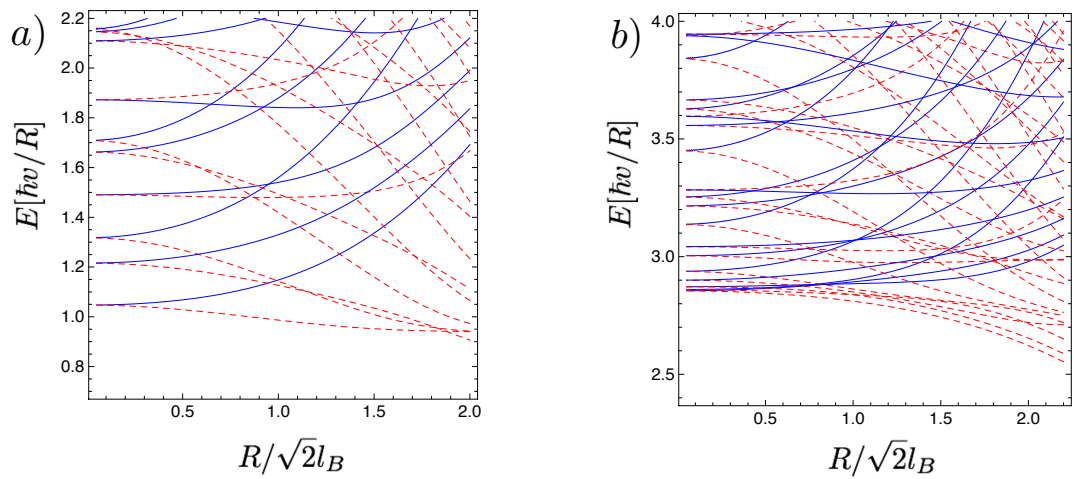}
\end{center}
\caption{{\bf Energy levels of bilayer quantum dots.} Energy levels in a relatively small bilayer quantum dot (radius $R=25 \, {\rm nm}$) as a function of the magnetic field. Parameters are as follows: $t_{\perp}=0.4 \, {\rm eV} =15.19 \hbar v/R$, $ U_0=1.52 \hbar v/R$ and $s=1$ (i.e. positive $B$-field). The solid and dashed lines are for different valleys. a) For $V=1.9 \hbar v/R$, b) for $V = 6 \hbar v/R$ (after Ref.~\cite{Reche2009}).}
\label{bilayerlevels1}
\end{figure}

\begin{figure}[h]
\vspace{0.4cm}
\begin{center}
\includegraphics[width=0.9\columnwidth]{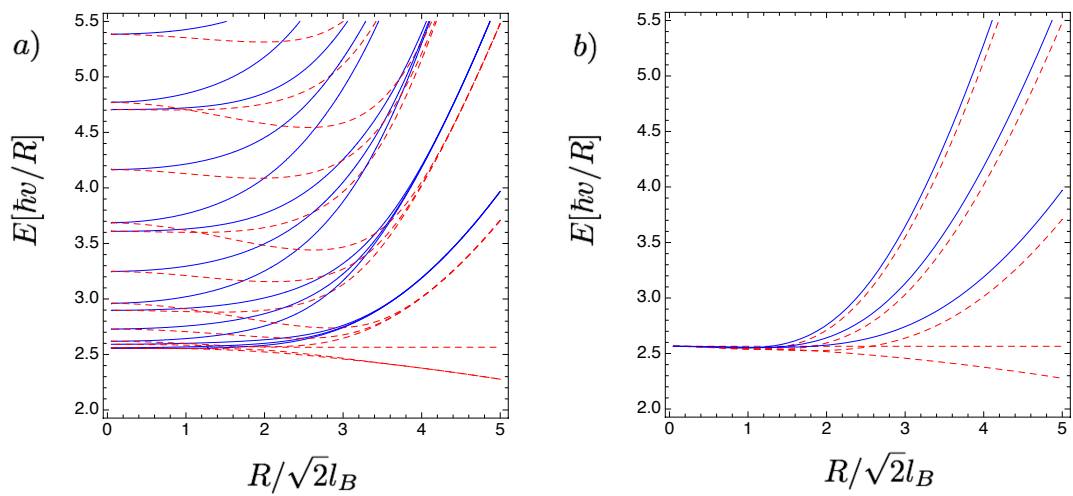}
\end{center}
\caption{{\bf Dot levels at high magnetic fields.} Merging of bilayer quantum dot levels to the bulk LLs as function of magnetic field for a relatively large bilayer QD with R=67.48 nm and $t_{\perp}=0.4 \, {\rm eV} =41  \hbar v/R$, $U_{0}=3.5 \hbar v/R$, $V=5.13 \hbar v/R$ for $m=0,\pm 1$ and $s=1$ (i.e. positive $B$-field). b) Bulk Landau Levels (LL) which are approached by the dot levels in a) almost perfectly at high fields in this parameter regime. Full lines are for $\tau=+1$ and dashed lines are for $\tau=-1$ (after Ref.~\cite{Reche2009}).}
\label{bilayerlevels2}
\end{figure}

The results for the bound state energies as a function of magnetic field are shown in Figs.~\ref{bilayerlevels1} and \ref{bilayerlevels2}. The most important result of our study can be seen in Fig.~\ref{bilayerlevels1} where we display the energy levels of a dot as a function of the magnetic field. At zero magnetic field, the degeneracy of the levels in the two valleys is clearly displayed. With increasing the magnetic field, the orbital degeneracy is lifted. The symmetry of the levels is analogous to the case of the single layer discussed above. The states that are degenerate at zero field are related by time-reversal symmetry which means that they correspond to opposite values of angular momentum $\pm m$ in different valleys. The typical eTRS of $\pm m$ within one valley is already broken by a finite "mass term", similar to the case of a disc in single-layer graphene discussed in Subsection \ref{singledisc}.

An important feature of the bilayer as opposed to the single layer is the unconventional Mexican hat-like dispersion relation near the band edge. This is most apparent for a large value of the bias field $V$, the case shown in Fig.~\ref{bilayerlevels1}b). It is clear that there are many closely spaced levels near the band edge. This is a feature of the enhanced density of states near this particular energy \cite{Nilss2007}.

For a large quantum dot, it is also possible to reach the regime where the dot levels are described by the Landau levels. This feature is seen in Fig.~\ref{bilayerlevels2}a) where we display the bound states for $m=0,m=\pm 1$ for large magnetic fields. Note that the dot levels tend to approach the bulk Landau levels displayed in Fig.~\ref{bilayerlevels2}b).

%%%%%%%%%%%%%%%%%%%%%%%%%
%%%%%%%%%%%%%%%%%%%%%%%%%
%This is a direct consequence of the mass-term $\tau\Delta\sigma_z$ in the dot Hamiltonian.
%Indeed, the symmetry operation ${\tilde T}=i\sigma_{y}{\cal C}$ with ${\cal C}$ the operation of complex conjugation, which transforms ${\bf p}%\rightarrow-{\bf p}$ and $\sigma\rightarrow -\sigma$ does not commute with the Hamiltonian in a single valley $[H_{\tau},{\tilde T}]\neq 0$.
%However, if both signs of $j$ were included, one would observe that the valley
%degeneracy was not broken at $B=0$, since the true time-reversal symmetry is not broken.
%%%%%%%%%%%%%%%%%%%%%%%%%%%%%%%%%%%%%%%%%%%%%%%%%
%%%%%%%%%%%%%%%%%%%%%%%%%%%%%%%%%%%%%%%%%%%%%%%%%%
\subsection{Symmetries and breaking of orbital degeneracy in graphene quantum dots}
\label{symmetries}
%%%%%%%%%%%%%%%%%%%%%%%%%%%%%%%%%%%%%%%%%%%%%%%%
In this subsection, we discuss the symmetries of graphene and their breaking by a smooth boundary which is the case for
quantum dots created by electrical gates, see Subsections \ref{singledisc} and \ref{bilayerdisc}. Here, we specialize to the case of single-layer graphene (similar arguments hold for the bilayer). The main point of Subsections \ref{singledisc} and \ref{bilayerdisc} has been to demonstrate that the valley-degeneracy can be lifted without mixing the valleys. In the following, we provide the reason using general symmetry arguments. The Hamiltonian Eq.~(\ref{htau}) breaks two symmetries present in pure graphene (i) the effective time-reversal symmetry ${\bf p}\rightarrow -{\bf p}$ and $\sigma\rightarrow -\sigma$, (ii) inversion symmetry ${\bf x}\rightarrow -{\bf x}$ of the graphene lattice. Both symmetries (i) and (ii) would lead to degeneracies in the spectrum---(i) within the same valley, (ii) within different valleys. In the case, where the valleys remain uncoupled by the boundary of the dot, there is another fundamental symmetry that leads to a degeneracy of states in different valleys, which is time-reversal symmetry ${\cal T}=-(\tau_{y}\otimes \sigma_{y}){\cal C}$ \cite{Beena2008}. However, time-reversal is easily broken by applying a magnetic field. Effective time-reversal symmetry (eTRS) ${\widetilde {\cal T}}=i\sigma_y {\cal C}$ and inversion symmetry ${\cal I}=-(\tau_{x}\otimes \sigma_z){\cal R}$ are both broken by the mass term $\Delta \tau_{z}\otimes \sigma_{z}$, where ${\cal C}$ and ${\cal R}$ denote complex conjugation and space inversion, respectively.

The magnetic field in $z$-direction alone breaks $ {\cal T}$ and ${\widetilde {\cal T}}$, but not $ {\cal I}$. Only the combination of a mass-term (generated by the substrate in single-layer graphene and the voltage $V$ between the two layers in bilayer graphene) and the magnetic field breaks all the orbital degeneracies. In Subsections \ref{singledisc} and \ref{bilayerdisc}, we have provided two realistic systems where these symmetries are broken.

Additionally, we mention here that in single-layer graphene the symmetries ${\widetilde {\cal T}}$ and $ {\cal I}$ are generally broken by any termination of the graphene lattice that does not couple the two valleys. At the boundary, the spinor needs to satisfy the linear equation $\psi={\cal M}\psi$ where ${\cal M}$ is in general a $4\times4$ matrix of the form ${\cal M}={\bf \nu}\cdot\tau\otimes {\bf n}_{\perp}\cdot \sigma$ where ${\bf \nu}$ is a vector on the Bloch sphere and ${\bf n}_{\perp}$ is a vector in the plane tangential to the boundary \cite{Akhme2008}. One easily shows that $[{\cal M},{\widetilde {\cal T}}]\neq 0$ and $[{\cal M},{\cal I}]\neq0$ if the vector $\nu=\pm {\hat z}$ (i.e. if the valleys are not coupled). Then, it is sufficient to additionally break $ {\cal T}$ by an external magnetic field if one wants to lift the valley-degeneracy  \cite{Reche2007}.

%%%%%%%%%%%%%%%%%%%%%%%%%
%%%%%%%%%%%%%%%%%%%%%%%%%
\section{Spin qubits in graphene}
\label{spinqubits}
%%%%%%%%%%%%%%%%%%%%%%%%%
%%%%%%%%%%%%%%%%%%%%%%%%%
Since the pioneering proposal on spin qubits for solid state quantum information processing \cite{Loss1998}, astonishing experimental achievements have been made, particularly in GaAs-based quantum dots, see Ref.~\cite{Hanso2007} for a recent review. In these (up to now) most advanced spin qubit systems, two major sources of spin decoherence have been identified: (i) spin-orbit interaction in combination with electron-phonon coupling to lattice vibrations, and (ii) hyperfine interaction of the electron spin with the surrounding nuclear spins.

Generally speaking, the ideal qubit should be easy to manipulate and should couple rather little to its environment. Therefore, it is natural to think about new host materials where at least one of these two points can be optimized. We show below that spin qubits in graphene quantum dots gain on both sides. They have more flexibility on the manipulation side and are superior as far as spin decoherence is concerned as compared to more conventional semiconductors like GaAs.
%%%%%%%%%%%%%%%%%%%%%%%%%
%%%%%%%%%%%%%%%%%%%%%%%%%%%%%%%%%%%%%%%%%
%%%%%%%%%%%%%%%%%%%%%%%%%%%%%%%%%%%%%%%%%
\subsection{Manipulation of spin qubits in graphene quantum dots}
%%%%%%%%%%%%%%%%%%%%%%%%%%%%%%%%%%%%%%%%%
%%%%%%%%%%%%%%%%%%%%%%%%%%%%%%%%%%%%%%%%%
For universal quantum computing, single-qubit and two-qubit manipulations are necessary. Single qubit rotations of spin qubits are naturally done by electron spin resonance (ESR) -- as proposed in Ref.~\cite{Engel2001} and measured in Ref.~\cite{Koppe2006} -- or by electric-dipole-induced spin resonance (EDSR) -- as proposed in Ref.~\cite{Borha2006} and measured in Ref.~\cite{Nowac2007}. The Rabi frequency $f_{\rm Rabi}$ at which the qubit rotates, for instance, in the ESR experiment \cite{Koppe2006} is proportional to the electron spin $g$-factor $f_{\rm Rabi} = g \mu_B B_{\rm ac}/2h$ where $\mu_B$ is the Bohr magneton and $B_{\rm ac}$ the external oscillating magnetic field used to rotate the spin. Notably, the electron spin $g$-factor differs for different materials. In GaAs quantum dots, it has been measured to be $|g| < 0.43$ \cite{Hanso2003} whereas, in graphene quantum dots, it has recently been determined to be close to $|g| = 2$ \cite{Guetti2010}. Thus, it is possible to rotate the electron spin in graphene quantum dots using ESR about five times faster than in GaAs quantum dots using the same field strength of the external oscillating magnetic field. This is an important gain because all qubit manipulations need to be done fast to be able to implement fault-tolerant quantum computing \cite{Niels2000}.

Another important advantage of graphene spin qubits is related to the small band gap in graphene nanoribbons. (For a ribbon width of about 30nm, the band gap can be estimated to be of the order of 60 meV.) This fact yields additional flexibility for two-qubit operations. Two-qubit operations are usually done via the Heisenberg exchange interaction $H_{\rm ex} = J \mathbf{S}_1 \mathbf{\cdot} \mathbf{S}_2$ where $\mathbf{S}_i$ ($i=1,2$) denotes the spin of the coupled qubits \cite{Burka1999}. Within a Hubbard model approximation with $t \ll U$, where $t$ is the tunneling matrix element between the quantum dots and $U$ the onsite Coulomb energy, the singlet-triplet splitting $J$ can be estimated to be $J \approx 4t^2/U$. The value of $U$ is usually fixed by the device geometry and the electromagnetic environment of the quantum dot. For graphene quantum dots, typical values of $U$ are about 10 meV, see, for instance, Ref.~\cite{Stamp2009}. The tunneling matrix element $t$ can, however, be easily tuned by increasing or decreasing the overlap of the wave functions of the electrons in the two quantum dots. In graphene or better to say in small band gap semiconductors, this manipulation can be done in two distinct ways. Either through tunneling via conduction band states (i.e. {\it normal tunneling}) or through tunneling via valence band states (i.e. {\it Klein tunneling}). This has been predicted for graphene nanoribbons in Ref.~\cite{Trauz2007} and experimentally realized in carbon nanotube quantum dots in Ref.~\cite{Steel2009}.

%
%%%%%%%%%%%%%%%%%%%%%%%%%%%%%%%%%%%%%%%
%%%%%%%%%%%%%%%%%%%%%%%%%%%%%%%%%%%%%%%
%%%%%      FIGURE    LDC         %%%%%%
%%%%%%%%%%%%%%%%%%%%%%%%%%%%%%%%%%%%%%%
%%%%%%%%%%%%%%%%%%%%%%%%%%%%%%%%%%%%%%%
\begin{figure}
\vspace{0.5cm}
\begin{center}
\includegraphics[scale=0.65]{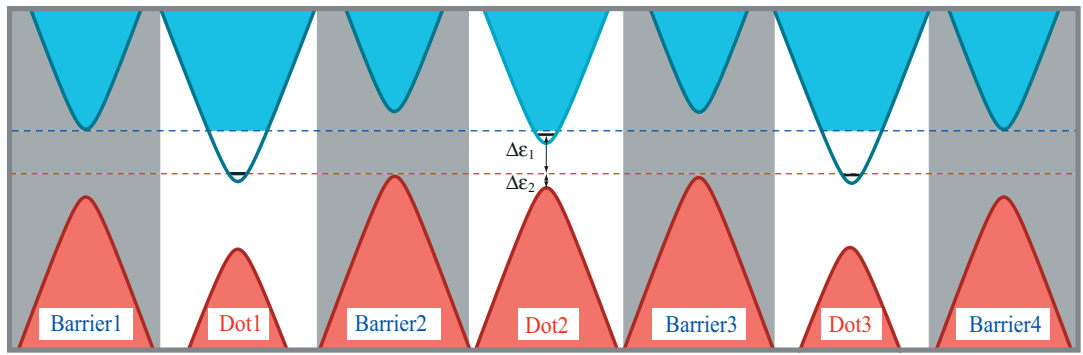}
\caption{\label{fig_LDC}
{\bf Long distance coupling of three graphene qubits.} The energy bands of three quantum dots (including barrier regions) are shown in which dot 1 and dot 3 are strongly coupled via cotunnelling
processes through the valence bands of barrier 2, barrier 3, and dot 2. (The valence bands are shown in red and the conduction bands in blue.) Importantly, the center dot 2 is decoupled by detuning.  The energy levels are chosen such
that $\Delta \varepsilon_2 \ll \Delta \varepsilon_1$ (after Ref.~\cite{Trauz2007}). }
\end{center}
\end{figure}

\begin{figure}
\vspace{0.5cm}
\begin{center}
\includegraphics[scale=0.62]{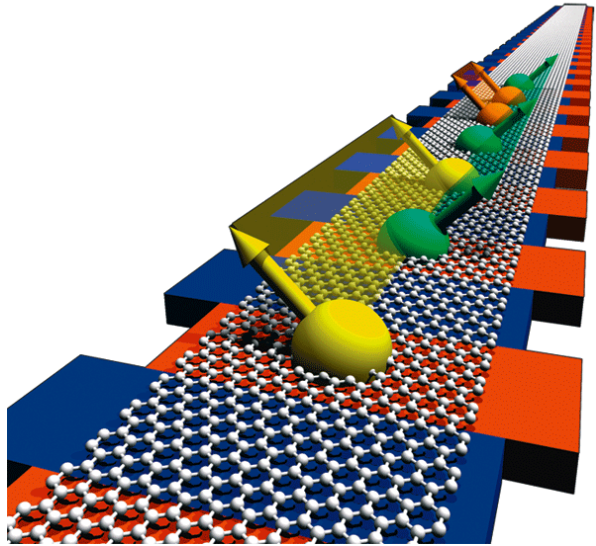}
\caption{\label{fig_piano}
{\bf Qubit piano.} Illustration of many spin qubits in a line hosted within a graphene nanoribbon. Quantum dots are red bars and barrier regions are blue bars. Different spin qubits that are strongly coupled to each other via Klein tunneling are marked with the same color (after Ref.~\cite{Falko2007}). }
\end{center}
\end{figure}

The most important physical consequence of this additional flexibility is the appearance of a new type of long-distance coupling between graphene spin qubits as illustrated in Fig.~\ref{fig_LDC}. By means of Klein tunneling, two distant qubits can be strongly coupled without touching the states of intermediate qubits that might be located between the two. Thus, a ribbon of graphene hosting many spin qubits in a line can be viewed as a qubit piano where any two of them can be coupled with leaving the states of the others unchanged, see Fig.~\ref{fig_piano} for a schematic illustration. Interestingly, this feature, i.e. the availability of non-local interactions, is important for quantum error correction since it raises the threshold for fault-tolerant quantum computing \cite{Svore2005}.

%%%%%%%%%%%%%%%%%%%%%%%%%%%%%%%%%%%%%%%
%%%%%%%%%%%%%%%%%%%%%%%%%%%%%%%%%%%%%%%
%%%%%      FIGURE    Piano       %%%%%%
%%%%%%%%%%%%%%%%%%%%%%%%%%%%%%%%%%%%%%%
%%%%%%%%%%%%%%%%%%%%%%%%%%%%%%%%%%%%%%%

In the previous section, we have given convincing arguments why the spin qubit manipulation in graphene has decisive advantages as compared to other commonly used host materials. In the next section, this advantage will be complemented by promising spin relaxation and dephasing properties of electrons and holes in graphene quantum dots.

\subsection{Spin relaxation and dephasing in graphene quantum dots}

Why can we expect stable spin qubits in graphene quantum dots? There is hope that spin relaxation and dephasing will be very weak in graphene for the following reasons: (i) Carbon is a light element with atomic number 6. Hence, its atomic spin-orbit interaction is weak as compared to heavier elements. However, such a statement should be taken with care because, in the solid state, spin-orbit coupling is often times dominated by bulk inversion or structure inversion asymmetry. Therefore, crystal structures of light elements can (under certain circumstances) exhibit rather strong spin-orbit coupling. Prime examples are carbon nanotubes where theory predicted a substantial spin-orbit coupling (a few hundred $\mu$eV) due to the curvature of the tube \cite{Ando2000,Huert2006,Bulae2008} which has been nicely confirmed in recent transport experiments on carbon nanotube quantum dots \cite{Kuemm2008}. Since the surface of graphene is less curved than that of carbon nanotubes, the spin-orbit coupling in graphene -- due to ripples -- should still be rather weak (roughly ten times less than the spin-orbit coupling due to curvature in carbon nanotubes \cite{Huert2006}). (ii) Carbon has two stable isotopes: $^{12}$C and $^{13}$C. The natural abundance is 99\% $^{12}$C and 1\% $^{13}$C. Since $^{12}$C has nuclear-spin $0$ and $^{13}$C has nuclear-spin $\frac{1}{2}$, the electron spin of the qubit can only interact with 1\% of the nuclei via hyperfine interaction. This ratio can even be further decreased because it is possible to artificially make $^{12}$C-enriched graphene.

\subsubsection{Spin relaxation due to spin-orbit interaction and electron-phonon coupling in graphene quantum dots.}

In the long-wavelength limit, the spin-orbit interaction in graphene can be written in the following form \cite{Kane2005,Min2006}
\begin{equation}
H_{\rm SOI} = \Delta_I \tau \sigma_z s_s + \Delta_R (\tau \sigma_x s_y - \sigma_y s_x) ,
\end{equation}
where $s_i$ denotes the Pauli matrix acting on the real spin, $\tau=\pm$ the valley degree of freedom, and $\sigma_i$ is the Pauli matrix associated with the pseudospin due to the two sublattices of graphene. The first term (proportional to $\Delta_I$) has been coined intrinsic spin-orbit interaction and is due to graphene's honeycomb lattice. The second term (proportional to $\Delta_R$) is of Rashba-type meaning that it is only present if the mirror symmetry about the plane is broken. This can be done by a perpendicular electric field, the interaction with the substrate, or ripples.

The electron-phonon coupling in graphene is characterized by two mechanisms called deformation potential mechanism (DPM) -- characterized by $V_1$ below -- and bond-length change mechanism (BLCM) -- characterized by $V_2$ below \cite{Suzuu2002,Ando2005,Maria2008}. The DPM is diagonal in pseudo-spin space whereas the BLCM is off-diagonal in pseudo-spin space. Thus, the electron-phonon coupling Hamiltonian can be written as
\begin{equation} \label{hepc}
H_{\rm EPC} = \left( \begin{array}{cc} V_1 & V_2 \\ V_2^* & V_1 \end{array} \right) + \; {\rm H.c.} \; .
\end{equation}
For spin relaxation in the low energy regime only acoustic phonons matter. There are three different acoustic phonons: (i) longitudinal acoustic (LA), (ii) transversal acoustic (TA), and (iii) transversal out-of-plane (ZA) phonons. Note that both LA and TA phonons have a linear dispersion whereas ZA phonons obey a quadratic dispersion. The specific form of $V_1$ and $V_2$ in Eq.~(\ref{hepc}) depends significantly on the type of phonons considered, see \cite{Bulae2008,Suzuu2002,Ando2005,Maria2008,Marti2003,Goupa2005,Struc2010} for concrete examples both for graphene and carbon nanotubes.

In a recent article, Struck and Burkard have shown that to lowest order in the spin orbit interaction as well as the electron-phonon coupling in flat graphene only the LA phonons matter for spin relaxation via the DPM while LA and TA phonons matter for spin relaxation via the BLCM \cite{Struc2010}. They find spin relaxation times $T_1$ between $10^{-1}$s and $10^{-5}$s depending on the applied external magnetic field $B$ that sets the level spacing between the two Kramers partners that form the qubit. This result should be compared to the corresponding analysis in GaAs \cite{Khaet2001,Golov2004} as well as carbon nanotube quantum dots \cite{Bulae2008}. Interestingly, the predicted numbers for $T_1$ are all in a similar range for different types of quantum dots but the $B$-dependence of the spin relaxation rate is totally different in different host materials. This is due to different symmetries and processes that dominate the relaxation. It confirms that a deep (microscopic) understanding of spin relaxation and dephasing in graphene quantum dots is crucial for an optimized qubit operation. Up to date, the spin decoherence time $T_2$ due to spin-orbit interaction in combination with electron-phonon coupling has not been calculated for graphene quantum dots. For GaAs \cite{Golov2004} and carbon nanotube quantum dots \cite{Bulae2008}, it has been shown that $T_2 = 2T_1$ to leading order perturbation theory in spin-orbit interaction and electron-phonon coupling.

All results of this section correspond to single layer graphene. In practice, it might turn out that spin qubits in bilayer graphene quantum dots are more stable than their single-layer counter parts. Rather little is known about the spin-orbit interaction \cite{Gelde2009,Guine2010,Liu2010} and electron-phonon coupling in bilayer graphene. Up to now, nothing is known about spin relaxation and dephasing in bilayer graphene spin qubits. We expect this to be a very active area of future research.

\subsubsection{Spin decoherence due to hyperfine interaction in graphene quantum dots.}

First investigations of hyperfine interaction in carbon nanosystems have been done on fullerenes -- based on NMR theories and experiments \cite{Penni1996}. Later on, ab initio calculations of the hyperfine interaction in small graphene nanoflakes have been carried out by Yazyev \cite{Yazye2008}. However, only very recently analytical calculations of the nuclear-spin interactions of electrons confined to graphene and carbon nanotube quantum dots have been presented \cite{Fisch2009}. The graphene part of the latter results will be reviewed in this section. Interestingly, pioneering experiments on nuclear-spin interaction and electron-spin dynamics in carbon nanotube quantum dots report an unexpectedly strong hyperfine interaction of order 100 $\mu$eV in $^{13}$C-enriched nanotubes \cite{Churc2009a,Churc2009b}. Unfortunately, similar data in graphene quantum dots does not exist up to now. Therefore, further experimental as well as theoretical work is needed to fully understand the role of
nuclear-spin interaction on spin relaxation in carbon-based quantum dots.

In principle, there are three terms that couple the spin of the confined electron to the nuclear
spins: (i) the Fermi contact interaction, (ii) the anisotropic hyperfine
interaction, and (iii) the coupling of electron orbital angular momentum to the nuclear spins.
These interactions are represented by the Hamiltonians \cite{Stone1975}
\begin{eqnarray}
\label{ham:contact}
h_1^k &=& \frac{\mu_0}{4 \pi} \> \frac{8 \pi}{3} \> \gamma_S \gamma_{j_k} \>
\delta(\mathbf{r}_{k}) \> \mathbf{S} \cdot \mathbf{I}_k,\\
\label{ham:anisotropic}
h_2^k &=& \frac{\mu_0}{4 \pi} \> \gamma_S \gamma_{j_k} \> \frac{3 (\mathbf{n}_k \cdot
\mathbf{S}) (\mathbf{n}_k \cdot \mathbf{I}_k) - \mathbf{S} \cdot \mathbf{I}_k}
{r_{k}^3 (1+d/r_{k})},\\
\label{ham:angular}
h_3^k &=& \frac{\mu_0}{4 \pi} \> \gamma_S \gamma_{j_k} \> \frac{\mathbf{L}_k \cdot
\mathbf{I}_k} {r_{k}^3 (1+d/r_{k})},
\end{eqnarray}
respectively, where $\gamma_S=2\mu_B$, $\gamma_{j_k}=g_{j_k} \mu_N$, $\mu_B$ is the Bohr magneton,
$g_{j_k}$ is the nuclear g-factor of isotopic species $j_k$,
$\mu_N$ is the nuclear magneton, $\mu_0$ is the vacuum permeability,
$\mathbf{r}_{k} = \mathbf{r} - \mathbf{R}_k$
is the electron-spin position operator relative to the nucleus,
$d \simeq Z \times 1.5 \times 10^{-15} \, \mathrm{m}$ is a length of nuclear dimensions,
$Z$ is the charge of the nucleus, and $\mathbf{n}_k = \mathbf{r}_{k}/ r_{k}$.
$\mathbf{S}$ and $\mathbf{L}_k = \mathbf{r}_k \times \mathbf{p}$ denote the spin and
orbital angular-momentum operators (with respect to the $k^{\mathrm{th}}$ nucleus)
of the electron, respectively. The cutoff $1+d/r_k$ avoids unphysical divergences from expectation values of the Hamiltonians $h_2^k$ and $h_3^k$.

In order to determine an effective spin Hamiltonian, one first needs to calculate matrix elements of $h^k_{1}$, $h^k_{2}$, and $h^k_{3}$ with respect to the electron states $|\Psi_\sigma \rangle = |\Phi_\sigma;u_\sigma \rangle |\sigma \rangle$ where $\langle \mathbf{r} | \Phi_\sigma;u_\sigma \rangle = \Phi_\sigma(\mathbf{r}) u_\sigma(\mathbf{r})$. Here, $\Phi_\sigma(\mathbf{r})$ is the envelope function and $u_\sigma(\mathbf{r})$ the Bloch amplitude which has the periodicity and, thus, knows about the symmetry of the honeycomb lattice. In the following, the envelope function is not further specified and the Bloch amplitude is approximated by hydrogenic orbitals, see Ref.~\cite{Fisch2009} for more details. In flat graphene, the electron states are $p$-type, whereas in nanotubes and curved graphene they are $sp$-hybridized. We restrict ourselves to flat graphene in the remainder of this section. Then, the matrix elements with respect to $h^k_{1}$ and $h^k_{3}$ vanish, and only $\langle \Psi_\sigma | h_2^k | \Psi_{\sigma'} \rangle$ is finite. It gives rise to the following effective spin Hamiltonian
\begin{equation}
H_{\rm HFI} = \sum_k \left( A_k^x S^x I^x_k + A_k^y S^y I^y_k + A_k^z S^z I^z_k \right)
\end{equation}
with coupling constants $A_k^j = A_j v_0 |\Phi_\sigma(\mathbf{r_k})|^2$ (where $v_0$ is the area of a primitive unit cell). Interestingly, the hyperfine interaction is asymmetric in space with \cite{Fisch2009}
\begin{equation}
\label{hfi}
A_x = A_y = -\frac{A_z}{2} =
-\frac{\mu_0 \gamma_S \gamma_{^{13}\mathrm{C}} Z_{\mathrm{eff}}^3}{240 \pi a_0^3} \approx -0.3 \mu{\rm eV} ,
\end{equation}
where the $z$ direction is perpendicular to the graphene plane. In Eq.~({\ref{hfi}), $Z_{\mathrm{eff}}$ labels the effective screened nuclear charge and $a_0$ is the Bohr radius. It has recently been shown, in the context of transport calculations on carbon-based quantum dots in the so-called Pauli blockade regime, that the hyperfine Hamiltonian gives rise to a coupling between the valley degree of freedom of the electron and the nuclear spins \cite{Palyi2009}. Such a coupling is not important for the interaction between the electron spin and the nuclear spins discussed here.

We now address the spin decoherence properties of a single electron spin in a graphene quantum dot coupled to a bath of nuclear spins in the situation where a ``large'' magnetic field $B_z$ is applied perpendicular to the graphene plane. Specifically, this means that $b=g\mu_B B_z \gg A_j$ which in fact corresponds to a rather moderate magnetic field of about 5 mT. Under these circumstances, the main source of decoherence is pure dephasing due to nuclear-field fluctuations along the spin quantization axis in $z$-direction governed by the Hamiltonian
\begin{equation}
H=(b+h_z)S^z
\end{equation}
with $h_z = \sum_k A_k^z I_k^z$. The total number of nuclei can be written as $N=N_{12}+N_{13}$ where $N_{12}$ is the number of $^{12}$C nuclei and $N_{13}$ is the number of $^{13}$C nuclei. For $N_{13} \gg 1$ we can use the central limit theorem (compare with Ref.~\cite{Coish2004}) which yields the following Gaussian dynamics
for the transverse electron spin
\begin{equation}
\label{spin-dynamics}
 \langle S^+ \rangle_t = \langle S^+ \rangle_0 \> e^{-t^2/\tilde{\tau}_c^2},
\end{equation}
where $S^+ = S^x + iS^y$. The latter equation is written in the rotating frame with
rotation frequency $(\omega+p \eta A_z/2)/\hbar$, where $\omega = b-b_N$ with the nuclear
Zeeman energy $b_N = g_N \mu_N B_z$. The characteristic electron-spin decoherence time is given by \cite{Fisch2009}
\begin{equation}
\label{coherence-time-graphene}
\tilde{\tau}_c = \frac{2 \hbar}{\sqrt{1-p^2}} \> \frac{N}{\sqrt{N_{13}} A_z},
\end{equation}
where the polarization $p$ of the nuclear-spin system ($0 \leq p \leq 1$) is determined only by the distribution of
spin-up and spin-down $^{13}\mathrm{C}$ nuclei. In Fig.~\ref{fig:coherence_graphene}, we plot $\tilde{\tau}_c$ as a function of the relative $^{13}$C abundance showing very promising decoherence times for graphene quantum dots. The number for the decoherence time for the natural $^{13}$C abundance $\tilde{\tau_c} \gtrsim 80 \mu \mathrm{s}$ \cite{Fisch2009} should be compared to the predicted $5$ns for GaAs \cite{Coish2004} in the absence of nuclear-spin polarization. This is a gain of more that 4 orders of magnitude.

\begin{figure}[t]
\centering
\includegraphics[width=0.6 \columnwidth]{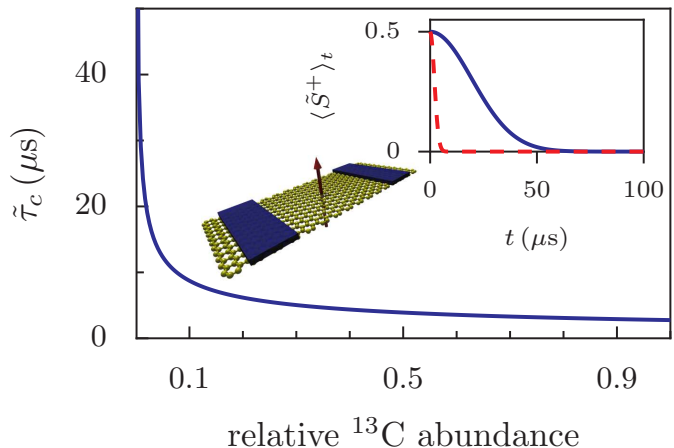}
\caption{{\bf Electron-spin decoherence time $\tilde{\tau}_c$ in graphene.} $\tilde{\tau}_c$ is plotted as a function of the
relative $^{13}\mathrm{C}$ abundance $N_{13}/N$, under the condition of a magnetic
field $B_z \gtrsim 5 \, \mathrm{mT}$ perpendicular to the graphene plane. We assume no nuclear-spin polarization and estimate the total number of nuclei to be $N=4 \times 10^5$.
Inset: Electron-spin dynamics in graphene with $^{13}\mathrm{C}$ abundances of
1\% (solid blue curve) and 99\% (dashed red curve), in the rotating frame (after Ref.~\cite{Fisch2009}).}
\label{fig:coherence_graphene}
\end{figure}

\section{Conclusions and Outlook}
We have reviewed the efforts towards the creation of quantum dots in graphene with the emphasis of using them as a host for spin qubits. Graphene has great potential for being an ideal candidate for spin qubits due to its low intrinsic spin-orbit coupling and the sparse amount of nuclear spins. We discussed theoretically in detail bound states in gate-tunable graphene quantum dots realized in graphene nanoribbons and in gapped single-layer and bilayer graphene. In contrast to quantum dots realized in edged graphene flakes, gate-tunable quantum dots are defined electrostatically and not by the physical edge of a graphene sample. This allows to controllably break the valley degeneracy, a prerequisite for spin-based quantum computing, e.g. by using a magnetic field. We have also discussed quantum manipulation of spin qubits in such dots as well as very recent theoretical studies on the consequences of spin-orbit interaction and hyperfine interaction with nuclei for spin-relaxation and spin-decoherence. Since theoretical as well as experimental efforts have so far focused mainly on single-layer graphene quantum dots, we believe that a great deal of interesting work will be devoted to bilayer graphene. Bilayer graphene is potentially superior to single-layer graphene due to the creation of a tunable bandgap by electric fields which allows for an all electrical control of graphene quantum dots.

We would like to thank C.W.J. Beenakker, Ya.M. Blanter, M. Borhani, D.V. Bulaev, G. Burkard, K. Ensslin, J. Fischer, D. Loss, A. Morpurgo J. Nilsson, A. Rycerz, V. Rychkov, C. Stampfer, E.V. Sukhorukov, M. Trif, J. Tworzyd{\l}o, and L.M.K. Vandersypen for collaborations and discussions on interesting research topics related to graphene quantum dots and spin qubits in these nanostructures. Financial support by the German DFG, in particular through the Emmy-Noether program (P.R.), is gratefully acknowledged.


\begin{thebibliography}{100}
\bibitem{remark1}
The relation between the Hamiltonians in Eqs.~(\ref{Dirac}) and (\ref{htau}) is a unitary transformation $U=i \sigma_y$ in the $K'$ valley.
\end{thebibliography}
\end{document}